\documentclass[manuscript]{acmart}
\AtBeginDocument{%
  }

\copyrightyear{2025}
\acmYear{2025}
\setcopyright{rightsretained}
\acmConference[IDC '25]{Interaction Design and Children}{June 23--26, 2025}{Reykjavik, Iceland}
\acmBooktitle{Interaction Design and Children (IDC '25), June 23--26, 2025, Reykjavik, Iceland}\acmDOI{10.1145/3713043.3731525}
\acmISBN{979-8-4007-1473-3/2025/06}
\acmISBN{978-1-4503-XXXX-X/2018/06}
\usepackage{multirow}

\begin{document}

\title[Building babyGPTs]{Building babyGPTs: Youth Engaging in Data Practices and Ethical Considerations through the Construction of Generative Language Models}

\author{Luis Morales-Navarro}
\email{luismn@upenn.edu}
\orcid{0000-0002-8777-2374}
\author{Daniel J. Noh}
\email{dnoh@upenn.edu}
\orcid{0009-0002-7219-1988}
\author{Yasmin B. Kafai}
\email{kafai@upenn.edu}
\orcid{0000-0003-4018-0491}
\affiliation{%
  \institution{University of Pennsylvania}
  \city{Philadelphia}
  \state{Pennsylvania}
  \country{USA}
}

\renewcommand{\shortauthors}{Morales-Navarro et al.}

\begin{abstract}
  As generative language models (GLMs) have gained popularity, youth are increasingly using them in their everyday lives. As such, most research has centered on supporting youth as users of GLM-powered systems. However, we know little of how to engage youth in the design of these models. Building on the rich legacy of child-computer interaction research that positions youth as designers of computing systems, we explore how to support young people in designing GLMs. Through a case study of three teenagers (ages 14–15) building a babyGPT screenplay generator, we illustrate how the team developed a model while engaging in artificial intelligence/machine learning-relevant data practices and addressing ethical issues. This paper contributes a case study that demonstrates the feasibility of engaging youth in building GLMs.
\end{abstract}



\maketitle

\section{Introduction}

While several efforts have studied how to engage youth in building artificial intelligence/machine learning (AI/ML) systems, these have focused on classification tasks ~\cite{druga2022family, kaspersen2023ai} with little attention given to generative tasks. Currently, most research conducted on youth’s use of applications powered by generative language models (GLMs) has centered on how youth build understandings of the systems’ functionalities through their everyday interactions ~\cite{solyst2024children, marx2024identifying}. Yet, we know little about youth’s understanding of GLMs when engaging in construction activities. In this paper, we investigate how youth engage in designing small GLMs by building on child-computer interaction's rich tradition of positioning young people as designers of computing applications ~\cite{harel1990software, druin1998design, iversen2017child}. In particular, we build on recent work on the role of construction in fostering computational empowerment by “engaging critically and curiously with the design of technology” ~\cite{dindler2020computational}. 

We conducted a five-day workshop with 35 youths (ages 14–15) in which they designed very small GLMs, which we call babyGPTs, using the nanoGPT framework \cite{KaparthyNano}. We analyzed their construction process to address the following question: How did youth engage with ML data practices and ethical considerations when ideating and building their own GLMs? Our analysis of a case study of one team showed that the three participants engaged in relevant data practices such as collecting data, controlling data quality, implementing a solution, and evaluating performance. At the same time, they considered ethical issues related to copyright, authorship, and attribution. This paper contributes a case study that demonstrates the feasibility of engaging youth in building GLMs. We discuss directions for future research on construction activities to support youth in designing GLMs.

\section{Background}

Over the past four years, with the release of easily accessible LLM-powered systems (such as ChatGPT and Gemini), researchers have turned their attention to youth and their use of GLMs. This work has investigated how to leverage GLMs in designing systems that support young people and their parents in reading \cite{dietz2024contextq, chen2024storysparkqa}, socio-emotional development ~\cite{li2024said, hu2024grow, hedderich2024piece}, and programming ~\cite{chen2024learning, druga2023scratch}. Other research has centered on the usage and understanding of LLM-powered systems, with studies on how young people use off-the-shelf generative AI/ML tools in creative tasks ~\cite{newman2024want, han2024teachers} and their everyday conceptions of commercial tools ~\cite{marx2024identifying, solyst2024children}. Several studies have investigated how interventions in which youth interact with GLMs, learn about their functionality, and explore their outputs may support learners in better understanding these systems and their implications \cite{ali2024constructing, solyst2024children}. These studies highlight how interventions may support youth in understanding GLMs. However, interventions have mostly focused on youth observing and evaluating outputs with no attention to the engagement of youth as designers of GLMs. To position youth as designers of GLMs, we turn our attention to the longstanding tradition in CCI of engaging youth as designers of computing applications.

Since the early days of CCI, a major concern has been involving youth as designers of computing applications \cite{papert1971twenty, harel1991children, kafai2012minds}. More recently, \citet{dindler2020computational} propose that engaging youth in both construction and deconstruction activities can support their development of computational empowerment. They conceptualize construction as activities in which learners’ “intentionality is materialized in technological artifacts through design and complex problem solving.” An important part of the construction process involves ideating and building \cite{iversen2018computational}. Ideating and building engages learners in coming up with ideas, transforming them into artifacts, and evaluating their relevance and performance. Furthermore, construction should also involve grappling with critical and ethical issues—considerations usually associated with deconstruction \cite{iivari2023computational}. 

Various efforts in CCI have engaged youth as designers of machine learning models. Most of these efforts have centered on classification tasks, where participants create small data sets that can be easily and quickly refined to train models and improve their performance \cite{arastoopour_irgens_characterizing_2022, druga_landscape_2022, hjorth2021naturallanguageprocesing4all, tseng_plushpal_2021, bilstrup2024ml}. \citet{vartiainen2021machine} argue that such an approach to ML fosters "data-driven reasoning and design," which entails considering dataset design decisions to explain the behaviors of machine learning systems. Recently, researchers have started to outline data practices that youth engage with when building models, including practices related to understanding a task, collecting data, understanding data, preparing data, implementing a solution, evaluating performance, deploying, and monitoring \cite{tseng2024co, olari2024data}. Several studies have investigated how youth build language models to classify words or sentences, highlighting ways in which youth can engage with natural language processing by connecting data-driven exploration to real-world applications \cite{hjorth2021naturallanguageprocesing4all, norouzi_lessons_2020, alvarez2022socially, jiang2023high, chao2023exploring}. 

In this paper, we focus on how participants conceived and built a very small GLM. To our knowledge, little research has been conducted on how youth and children engage in building generative models. One study by \citet{williams2019popbots} involved children in creating generative music by choosing emotional states with tempo and chord progression parameters and using an interface to feed training data to the system. Other efforts have involved discussing general adversarial networks (GANs) and distinguishing GAN-generated media from human media, as well as addressing the ethical implications of generative image models \cite{lee2021developing, ali2024constructing} without youth building datasets or training models. We build on these experiences to investigate how youth engage in the construction of GLMs. 

\section{Methods}

\subsection{Participants and Context}

We conducted a five-day workshop (90 minutes a day) with 35 ninth graders (ages 14–15) and two teachers at a public school in the United States. All participants had taken at least one year of programming classes at school. In this paper we focus on three teenagers: Dillpickle, Optimus, and Cyclops. Two of them identified as male and one as non-binary; all participants identified as White. All names in the paper are pseudonyms.\footnote{We asked all participants to decide on their own pseudonyms \cite{kivuva2024cultural}.}

\subsection{Workshop Activities}
During the workshop, participants explored GLMs by actively creating their own. Two STEM and computing teachers facilitated these activities, with support from a researcher. The first day of the workshop centered on having youth reflect on how they use GLMs in their everyday lives and their understanding of how these models function. The second day of the workshop included an introduction of key ideas about AI/ML, a game about probability and Markov chains, and a hands-on activity where youth created their own GLMs using Markov chains. The third day centered on building babyGPT datasets, and the fourth day on exploring babyGPT outputs. The last day of the workshop focused on the implications of GLMs. Below, we briefly describe the activities for days three and four, which are the focus of this paper. 

\paragraph{Day 3: Building a babyGPT Dataset}
On the third day, the teachers introduced LLMs and the role of probabilities in generating synthetic text in LLMs and neural networks at large. Here, the teachers highlighted the role that humans play in curating datasets intentionally, evaluating outputs to prevent harm, and identifying tasks in which non-deterministic outputs are appropriate to prevent misinformation and misuse of GLMs. Following, participants explored the outputs of Bhatia's \cite{Bhatia2023} babyGPT project for the New York Times.\footnote{Using Kaparthy's \cite{KaparthyNano} nanoGPT,  a rewrite and reimplementation of OpenAI’s GPT-2 to quickly train and finetune medium-sized GPTs, framework, \citet{Bhatia2023} created GLMs based on the works of Jane Austen, Shakespeare, and Moby-Dick.} The class discussed the differences in the size of datasets and computing power between models such as GPT 3.5 and Bathia’s babyGPTs. 

Teachers introduced the babyGPT activity, explaining that the youth would decide the type of GLM they would make, decide how they would create a dataset, and tokenize\footnote{Tokenization is the process of transforming text into a sequence of tokens, which can be words, characters, or subwords. This is necessary in preparing data for training; in tokenizing data, each token is assigned a unique integer used to create a vector representation that is then used in training a model.} their dataset using the provided script. While the youth brainstormed themes for their GLMs, the teachers invited them to reflect on the following questions: “What kind of data do you want to use in your model?” and “Is it okay to train models with data without permission from the people that created the data?” The youth were tasked with designing a small dataset containing 75,000 to 300,000 tokens. They then brainstormed potential prompts, which the teachers introduced as “seeds” for generating text. Finally, they submitted training orders, specifying how long they wanted their models to be trained for. The models were trained by a researcher after the workshop using Kaparthy's \cite{KaparthyNano} nanoGPT framework. All models were trained between 5 min (200 iterations) and 1 hour (2400 iterations).\footnote{While the youth’s computers had the infrastructure to train models (using M1 chips), for security reasons, school administrators blocked participant access to the terminal, preventing youth from training the models at school.}

\paragraph{Day 4: Exploring babyGPT Outputs}
The fourth day began with teachers discussing how dataset size, computing power, data quality, and model training duration influence model performance. Teachers then highlighted the differences between the dataset size of the babyGPT models (4,000–300,000 tokens) and GPT-4 (more than 1 trillion tokens \cite{yenduri2024gpt}), the duration of model training, and the computing power used to train them. This helped align the youth’s expectations with the capabilities of their models. On this day, youth played with their 1-hour models, looking at different outputs, comparing the 1-hour models to 5-minute or 30-minute models, looking at each other's models, and brainstorming how the models could be improved.

\subsection{Data Collection and Analysis}

We collected four primary sources of data: video recordings of youth engaging in the activities, images of artifacts the youth created (e.g., pictures of big papers), the actual files of the datasets, models, and model outputs. Due to the novelty of this project, we chose an exploratory descriptive case study \cite{yin2018case} as a method because it  allowed us to analyze how one group of participants ideated and built their babyGPT model. From all groups, we chose the group with the most complete data (e.g., filled-in worksheets, saved copies of their dataset, and attendance). We constructed a descriptive case study.\footnote{Descriptive case studies have been used in human-computer interaction (HCI) to illustrate the context-specific nature of participatory design research with teens \cite{duarte2018participatory}. Several single-case studies have been previously published in the IDC community \cite{davis2017haptic, bell2015learning, di2009interactive}, showing how exploratory work can lay the groundwork for future research.} This type of case study aims to depict a phenomenon in context rather than providing explanations or considering competing interpretations ~\cite{yin2018case}. In our analysis, we also relied on Olari's \cite{olari2024data} inventory of AI/ML data practices to track how the group engaged with these practices while building their babyGPTs. Two researchers watched and logged 6.5 hours of video recordings. We then organized the video logs using \citet{iversen2017child}'s
construction process as a descriptive framework and supplemented them with data from youth-designed artifacts.

\section{Findings}
When ideating and building their babyGPT projects, youth engaged with the following AI/ML data practices: collecting data, understanding the data, controlling data quality (data validity), preparing the data, and evaluating performance. They also engaged with ethical considerations related to copyright, authorship, and attribution. 

\paragraph{Building a Dataset}

When building the dataset, team members thought carefully about what screenplays to add to their dataset. Particularly, Cyclops suggested that they do not collect data from new Marvel movies, Dillpickle agreed, “100\%, every Marvel movie going up until Endgame.” The team agreed that, out of the 62 Marvel movies, they would only add scripts of their favorite ones. The team understood the importance of data quality and how their existing knowledge of the Marvel universe could help them make decisions about the data needed for their project. For example, they discussed whether to include Spiderman scripts in their dataset and, despite not being fans of the newest Spiderman film, ultimately decided to include them. Cyclops argued against including the Hulk, and Optimus argued for including Thor, the Avengers, and Deadpool.


The team simultaneously collected and prepared the data by searching for scripts while tokenizing them. To create a dataset, Cyclops proposed that they searched for movie scripts online
. He proceeded to copy a script of Iron Man 1 into a text file and ran the tokenization script, observing that the script had produced 30,000 tokens. He then found the script of Iron Man 2 and noted that it was only 15,000 tokens—significantly shorter than Iron Man I. Dillpickle warned Cyclops, “We’ll need a lot of text for it to work,” reminding him that the GLMs needed 70,000 to 300,000 tokens. Cyclops continued by adding scripts of Captain America, Avengers: Infinity War, and Avengers: Endgame, eventually reaching 80,050 tokens. Learners' engagement with data practices was not linear but rather iterative \cite{olari2024data}. In the collection and preparation of their dataset, the team exhibited the highly iterative nature of these data practices through an extended back and forth between looking for scripts and tokenizing their data.  

While preparing their data for training, team members shared different stances towards the ethical considerations of using screenplays to train a model. A researcher asked the team about their thoughts on training models with data without the permission of the data creators. Optimus argued that because screenplays “are copyrighted, we can’t use them.” Dillpickle suggested another idea, emphasizing the importance of attribution by noting that screenplays “are copyrighted, and we’re going to use the name of their characters and some of their dialogue. I think that the people that created those characters should get some credit.” Cyclops, who initially answered, “Sure, why not?” to the researcher’s question, later agreed with Dillpickle, stating that attribution is important for the “same reason we can’t just copy normally because it’s not your own work.” He then considered a hypothetical scenario: "What if the script the model creates is super good and then we sell it to Marvel for tons of money? They’re gonna ask how 15-year-olds come up with a script! But we didn't; we just used their scripts to train AI; now give me money.” In considering this hypothetical scenario, Cyclops framed generating screenplays using GLMs as inherently different from human authorship.

After designing a dataset, the youth filled out a model training request form with options to train for 5 minutes, 30 minutes, 1 hour, and 2 hours. The team chose to have their model trained for 5 minutes and 1 hour compare the differences. The models were trained by researchers overnight with 200 training iterations for the five-minute model and 2400 training iterations for the one-hour model.

\paragraph{Examining Outputs}

\begin{figure}
    \centering
    \includegraphics[width=0.8\linewidth]{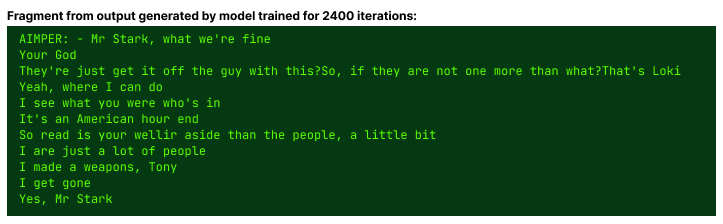}
    \caption{Fragment from output of the Marvel babyGPT model.}
    \label{fig:marvel}
    \Description[Screenshot of outputs for a marvel screenplay generator designed by Cyclops, Dillpickle, and Optimus.]{Screenshot of outputs for a marvel screenplay generator designed by Cyclops, Dillpickle, and Optimus. Text in the screenshot reads: AIMPER: - Mr Stark, what we're fine Your God They're just get it off the guy with this?So, if they are not one more than what?That's Loki Yeah, where I can do I see what you were who's in It's an American hour end So read is your wellir aside than the people, a little bit I are just a lot of people I made a weapons, Tony I get gone Yes, Mr Stark}
\end{figure}

As the team examined the outputs of their babyGPT (see Figure \ref{fig:marvel}), they demonstrated another data practice, evaluating performance. The next day in class, once their model was trained, the team explored the outputs of different prompts given to the trained babyGPT model. One teacher explained that prompts are like seeds that provide a starting point for the model to generate new text; that is, the model predicts the words that should follow the different prompts. The team came up with the following four prompts: “Write an ending where Iron Man doesn't die,” “Make Captain America evil,” “Let Thanos win,” and “Write about Tony.” They explored different outputs generated by the one-hour model for each prompt. Their observations about the outputs included how funny some of the sentences constructed by the model were, the lack of screenplay structure, and narrative issues. They compared the one-hour model to the five-minute model and to other models designed by their peers. Here, we focus on their exploration of the one-hour model trained for 2400 iterations. 

The team’s first reaction when exploring the outputs of the babyGPT model was laughter, finding many of the sentences created by the model amusing. The group began reading one of the outputs generated for the prompt “Make Captain America evil” in a very serious tone and then started laughing. To illustrate, halfway through the output, Dillpickle sternly read, “This is what I do, I'm gonna stop you,” a sentence that seemed like a plausible line from a Captain America screenplay. He continued, “I'll get just radioise me? - You're a little low, are you?- He was a radiation like you leave let's work.” Optimus laughed aloud, asking, “what is radiose?” As Dillpickle kept reading, Optimus laughed again, highlighting the absurdity of the phrase “he was a radiation” by repeating it twice. 

The team also noted how the outputs lacked the common structure of a screenplay, where character names are followed by their lines. Cyclops started reading an output for the prompt “Ironman does not die.” As he read the awkward sentence, "The only thing about to the world's working for you, by the plan, I want to have a plan,” Optimus exclaimed, “We have no idea who is saying this,” noting that the output text did not indicate which lines belonged to which character.

They also noted issues with the narrative flow, or lack thereof. As Cyclops finished reading an output that ended with the phrase “That's all, I'm always be a," he explained, “Here, this is bad,” arguing that the text did not make any sense. Similarly, as Optimus read an output to the prompt “Let Thanos win," he exclaimed, “What! It doesn’t even make sense!” A couple of minutes later, he expressed that “there’s no direction” in the script. He further explained, “It’s like if you drugged someone and forced them to write like a script; that’s what this sounds like—someone drunk trying to write this.” The process of working with the model helped the team recognize that meaning-making and storytelling require the thoughtful construction of a narrative arc, which extends beyond merely stringing related words together.

\section{Discussion}

We examined the feasibility of youth being designers—and not just users—of GLMs following a longstanding tradition of CCI research. Compared to other studies in which youth learned about the functionality of generative models and explored their outputs as users \cite{ali2024constructing, solyst2024children} engaging youth in construction activities made more transparent process of building models. Building babyGPTs supported youth to question the “aesthetic legitimacy” of outputs \cite{solyst2024children}, and voiced nuanced understandings of the functionality and ethics of GLMs. These findings indicate that positioning youth as designers of GLMs should be pursued further to promote development of AI/ML literacies.

In ideating and building models, which also involves evaluating their relevance and performance, the youth engaged in key data practices such as data quality control, data collection, data preparation, and the evaluation of the performance of their models. As \citet{olari2024data} note, ML data practices and pipelines are not linear, as “students may move back and forth between the [practices].” In collecting and preparing their dataset, the youth went back and forth between looking for data and tokenizing their data. Future work should provide opportunities for more iterative engagement with both data practices and the construction process by providing more time for youth to iteratively design, train, and evaluate their models. Ideating and building are only one part of the construction process; defining a design problem, doing research, and arguing and reflecting are equally important parts of the construction process that should be further investigated.  

Our findings suggest that constructing GLMs should be an integral part of efforts to foster AI/ML literacies. As \citet{dindler2020computational} argue, the “construction of digital technology requires students to understand technology as a material that can be shaped and molded,” which in the case of GLMs involves understanding how these models are not magical but created by humans that make decisions about data, learning algorithms, and ethical and environmental issues. Future designs could elaborate on similar activities, by having youth explore validation and training loss, adjust weights, or finetune pre-trained models. This would also require developing easy-to-use tools that can support novices in building their own GLMs. We hope that this work motivates further research to support youth in designing and learning from designing GLMs.  

\section{Selection and Participation of Children}

We recruited a class of ninth-grade youth and their teachers from a high school in a city located in the Northeastern United States. Youth participated in the study as part of their classroom activities. Parents received consent forms prior to the study, which included a brief explanation of the research, and youth assented to their participation. Research protocols and data collection methods were approved by the IRB board of the University of Pennsylvania.

\begin{acks}
The implementation, analysis, and writing of this paper were supported by National Science Foundation grants \#2414590 and \#2333469. Any opinions, findings, and conclusions or recommendations expressed in this paper are those of the authors and do not necessarily reflect the views of NSF or the University of Pennsylvania.  With regard to Aatish Bhatia, whose babyGPTs inspired this project.  
\end{acks}

\bibliographystyle{ACM-Reference-Format}
\bibliography{references}

\appendix

\end{document}